# Unitary recordings in freely-moving pulse weakly electric fish suggest spike timing encoding of electrosensory signals

Alejo Rodríguez-Cattáneo*, Ana Carolina Pereira[1]*, Pedro A. Aguilera, and Ángel A. Caputi

Departamento de Neurociencias Integrativas y Computacionales,

Instituto de Investigaciones Biológicas Clemente Estable,

Av. Italia 3318, CP 11600, Montevideo, Uruguay

Corresponding Author: Angel Caputi, caputiangel@gmail.com; acaputi@iibce.edu.uy;

phone (598) 24871616, fax (598) 24875548

---

[1] *ARC and ACP have equal merit in this work.

**Abstract**

*Evaluation of neural activity during natural behaviours is essential for understanding how the brain works. Here we show neuron-specific self-evoked firing patterns, modulated by an object's presence, at the electrosensory lobe neurons of freely moving <u>Gymnotus omarorum</u>. This novel preparation shows that electrosensory signals in these pulse-type weakly electric fish are not only encoded in the number of spikes per electric organ discharge (EOD), as is the case in wave-type electric fish, but also in the spike timing pattern after each EOD, as found in pulse-type Mormyroidea. Present data suggest that pulsant electrogenesis and spike timing coding of electrosensory signals developed concomitantly in the same species, and evolved convergently in African and American electric fish.*

**Main text**

Electric fish construct sensory images by "illuminating" their nearby surroundings with a self-generated electric field – characterized by either pulsating patterns or continuous sine-waves (Moller, 1995). Corresponding to these object polarization strategies, primary afferents either modify the strength of a burst of spikes (in the case of pulse discharges) or the probability of firing (in the case of wave discharges) when the local intensity of the self-generated field is affected by the presence of an object. The electrosensory lobes of wave and pulse Gymnotiformes where afferent project show a similar neuroarchitecture (reviewed in Bell and Maler, 2005, illustrated in Fig 1a) but neurons appear to respond differently to the EOD. In wave Gymnotiformes either the spike rate of single neurons or the synchronous increment in rate within a neuronal sub-population encode the amplitude modulations of the local signals (reviewed in Krahe and Maler 2014; Clarke et al., 2015). In pulse Gymnotiformes, field potentials continuously recorded, and current source density analysis and unit recordings in decerebrated fish show clear post-EOD patterns (Pereira et al., 2005; Pereira et al., 2014). This raises the question on whether spike timing of different unit types carry information regarding nearby objects. To answer this question, unitary neuronal recordings from the electrosensory lobe of freely moving pulse gymnotiform fish were utilized. These recordings were carried out in 4 *Gymnotus omarorum* (12-15 cm length, unknown sex) under the protocol 001/003/2011 of the animal care committee of our Institute. All surgical and potentially-painful procedures were performed in anaesthetized fish (confirmed by lack of response to noxious stimuli). Lidocaine gel was applied to the surgical area to avoid cutaneous and bone-localized pain. At the end of the experiment, fish were euthanized (MS222, 1 %).

For chronic electrode implantation, a portion of the scalp was removed, and a pair of insulated nichrome wires (50 $\mu$m diameter, twisted in a double-helix fashion with tips exposed and separated by approximately 100 $\mu$m) was inserted through a 150 $\mu$m opening, positioned at the polymorphic layer of the centro-medial map of the electrosensory lobe (Pereira et al., 2005, Fig 1a), and cemented to the skull. The spikes and the far field EOD were recorded with differential amplifiers (1800, AM-systems, band pass: 300-3000 and 10-10000 Hz; gains: x10000 and x100, respectively), digitized, stored, and units were later sorted using commercial programs (Experimenter Datawave technologies). After 3 hs at room temperature (20°C), and a water conductivity of 100 $\mu$S/cm, the EOD rate returned to the previous baseline and intra-cerebral electrodes recorded field potentials and unitary activity characteristics of the polymorphic layer of the electrosensory lobe (Pereira et al., 2014).

Four units were recorded in the absence of objects in the first implanted fish, and 6 were clearly sorted in the other 3 fish. Sorting was performed by a) selecting epochs of 2.5 ms around the threshold crossing time stamps for each sharp deflection of the signal, and b) classifying them as 4 or 5-dimensional clusters according peak-to-peak heights, peak amplitudes, and peak timings.

Spike timing showed a non-uniform probability distribution following the EOD. All spike patterns exhibited the presence of a silence between 7 to 10 ms, and 1 to 3 modes found at ca. 5, 12, or >23 ms after the positive peak of the EOD (as in Pereira et al., 2014). Post-EOD spike histogram peaks were differently modulated, corresponding to the unit and sensory context. The best unit in each of the last 3 fish was selected to explore different effects (Fig 1. b to d: conductivity, plastic vs. metal cubes 2 cm side; e to g: movement; and h to j: tube hiding behaviour).

A static metallic cube before the skin either caused increases or decreases in the firing rate (Fig 1b-c). These increments may represent those that led the functional classification of electrosensory neurons as "centre-on" and "centre-off" in wave fish (Clarke et al., 2015). Strikingly distinct from wave fish, *G. omarorum* neurons showed changes in the post-EOD firing pattern in addition to the change in the number of spikes per EOD.

For the "centre-on" unit (Fig 1b) plastic and metallic objects caused opposite effects in the firing rate. However they did not provoke simple mirror-image changes, but drastic ones in the post-EOD pattern. Histograms constructed from samples exhibiting the same number of spikes were significantly different (Fig 1. b). Note the large increase of the peak at 12 ms after the metallic object is introduced, and the small adaptation effect.

Two clearly separable "centre off" units were explored in greater detail in two other fish. These cells show clear adaptation, and in one case, seemingly paradoxical responses. Their responses to the EOD show maximal increments when the object was approximated and slowly adapted towards a resting value even when fish remained stationary in front of it (Fig 1c).

One unit was also explored when the metal cube was manually moved back and forth, parallel to the skin at a distance of 1-2 mm. Aiming to stimulate with a "texture effect" previously described (Caputi et al., 2011) the cube face against the skin was carved with a saw-tooth profile. Back and forth movements of the cube caused strong modulations in spike firing rate and spike timing pattern following the EOD (Fig 1d). Histograms corresponding to object absence (Fig. 1f) and object movement (Fig 1g) were significantly different, and paradoxically for a "centre off" neuron, a moving metal object caused an increase in its firing rate – possibly due to their phasic responsiveness (compare the number of EODs necessary to recruit the same number of spikes; Fig 1. d, gray rectangles).

The other unit having a "Centre off" receptive field with receptive field at the mouth commissure was recorded when the fish took refuge inside a plastic tube (5 cm diameter, 10 cm long, with a slit for passing the wires). Figure 1 illustrate three stable conditions indicated at the insets. It was previously shown that when the fish body was completely inside the tube with the exception of its head ("head sticking out"; Fig. 1h), the local signal at the receptive field is increased respect to the control (Fig. 1i) but when the fish head is at the middle of the tube the local signal is reduced ("head inside"; Fig. 1j; Pereira et al., 2005). The "head sticking out" position (i.e. the most commonly observed during natural behavior, which also increases the signal at the receptive field centre) caused a global reduction in spike rate with a small effect on the early and a virtual disappearance of the late modal peak observed in the control condition. The "head inside" position (i.e. reduction of the stimulus at the receptive field centre) caused a sharp increase of the early modal peak, and a shift of the second mode.

In conclusion, this letter introduces the development of a novel technique utilized to perform unitary recordings in freely moving electric fish. This technique (which has in parallel developed by the group of Maler, Fotowat et al. 2019), once combined with electric image modelling (reviewed in Caputi and Budelli, 2006), will allow a fuller understanding of the active electric sense. Although further experiments should be used to pinpoint neurons phenotypes (to improve our understanding of the electrosensory lobe network), here we show that the post-EOD firing patterns characteristic of *G. omarorum* "centre off" units contain additional information on the electrosensory input that is not contained in the average number of spikes per EOD (e.g. spike rate). Finally, comparative analysis suggests a daring hypothesis: spike timing encoding of electrosensory signals have evolved concomitantly with the ability of explore the environment using pulsatile discharges in a convergent way in African and American species. In fact, the African wave fish *Gymnarchus niloticus* shows a rate code and pulse Mormyroidea show neuron-specific post-EOD firing patterns, both modulated by electrosensory stimuli (reviewed in Kawasaki, 2005).

**Author contributions:** design and first draft AAC; experiments and analysis: AAC, ARC, ACP, PAA.

**Acknowledgements:** Authors thank to Dr. J. Waddell for comments and English edition. Partially supported by ANII (PhD fellowships to ARC and ACP) and UDELAR (fellowship to ARC), Uruguay.

**References**

1. Bell, C.C. & Maler, L. (2005). Central neuroanatomy of electrosensory systems in fish. In *Electroreception* (pp. 68-111). Springer, New York, NY.

**2.** Caputi, A. & Budelli, R. (2006). Peripheral electrosensory imaging by weakly electric fish. *Journal of Comparative Physiology [A]* **192,** 587–600.

3. Caputi, A. A., Aguilera, P. A., & Pereira, A. C. (2011). Active electric imaging: body-object interplay and object's "electric texture". *PloS one*, **6(8),** e22793.

4. Clarke, S.E., Longtin, A., & Maler, L. (2015). Contrast coding in the electrosensory system: parallels with visual computation. *Nature Reviews Neuroscience*, **16(12),** 733

5. Fotowat H., Lee C, Jaeyoon Jun J, Maler L (2019) Neural activity in a hippocampus-like region of the teleost pallium is associated with active sensing and navigation. eLife **8,** e44119.

6. Kawasaki M. (2005) Physiology of tuberous electrosensory systems. In: *Electroreception*. (Bullock TH et al., eds.) Springer, New York, NY, p. 154-194.

7. Krahe, R. & Maler, L. (2014). Neural maps in the electrosensory system of weakly electric fish. *Current opinion in neurobiology* **24,** 13–21.

8. Moller, P. (1995). *Electric fishes: history and behavior* (Vol. 17). Chapman & Hall. NY

9. Pereira, A. C., Centurión, V. & Caputi, A. A. (2005). Contextual effects of small environments on the electric images of objects and their brain evoked responses in weakly electric fish. *Journal of Experimental Biology* **208,** 961–972.

10. Pereira, A.C., Rodriguez-Cattaneo, A., Caputi, A.A. (2014). The slow pathway in the electrosensory lobe of *Gymnotus omarorum*: Field potentials and unitary activity. *Journal of Physiology-Paris* **108,** 71–83.

11. Press, W. H., Teukolsky, S. A., Vetterling, W. T., Flannery, B. P. (1996). *Numerical recipes in C* (Vol. 2). Cambridge: Cambridge university press.

**Figure Caption:**

**Chronic recordings in *Gymnotus omarorum*.** a) Traces: EOD (top) and the electrosensory lobe (bottom) recordings; inset: overlapped waveforms of sorted spikes. Histology indicates the recording place (CM: centromedial, ML: mediolateral, L: lateral, maps, EGP: eminentia granularis posterior, RN:relay nucleus). Responses of a "centre on" (b) and a "centre off" (c) neurons when a 2 cm side cube was placed in front of the receptive field (shadow regions of the raster). Note that opposite but not "mirror-image effects" of equally shaped plastic (before the arrowhead pointing up) and metal (after the arrowhead pointing down) cubes on the "center on" neuron. Spike timing distributions evoked by plastic and metal cube showed a significant difference ($\chi^2$=221, DF=49, p<0.001, 200 spikes taken form left and right boxes in each histogram, Press et al, 1996 ). d) The same unit as illustrated in c, shows sharp variations of the raster responses to the EOD in the presence of "va-et-vien" object's movements. Gray bands encompassing 200 spikes during movements (d) and object absent (e) indicate the data used to construct significantly different histograms f and g ($\chi^2$=91, p<0.001, DF 24). The firing pattern of another "center off" neuron during refuge behavior: h) "head sticking out" condition (local EOD increased); i) control; j) "head fully inside" condition (local EOD reduced). Histograms constructed with 300 spikes each were significantly different ($\chi^2$ tests, DF=49, largest p<0.003 equivalent to 0.01 after Bonferroni's correction).

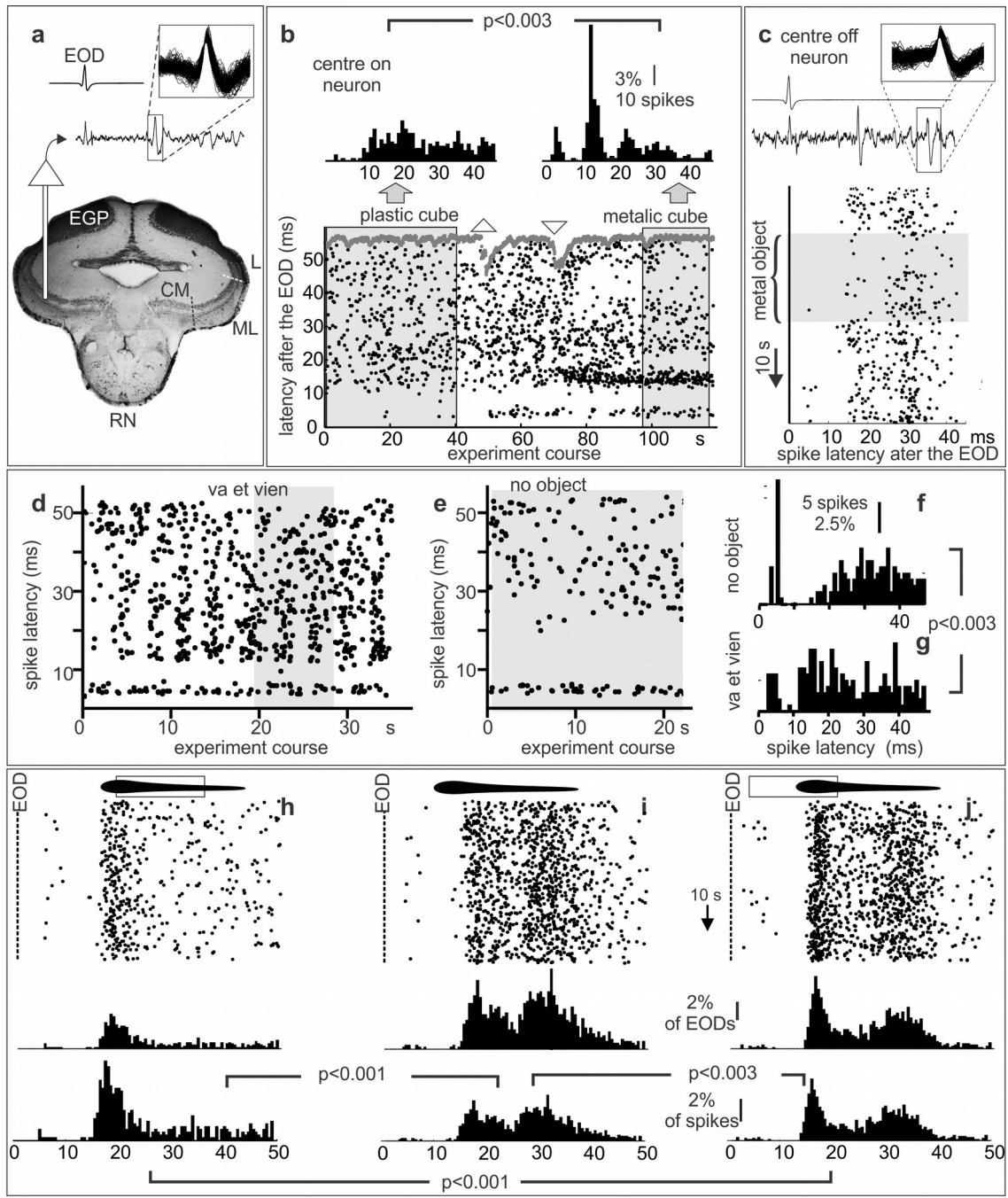

**Chronic recordings in *Gymnotus omarorum*.** a) Traces: EOD (top) and the electrosensory lobe (bottom) recordings; inset: overlapedd waveforms of sorted spikes. Histology indicates the recording place (CM: centromedial, ML: mediolateral, L: lateral, maps, EGP: eminentia granularis posterior, RN:relay nucleus). Responses of a "centre on" (b) and a "centre off" (c) neurons when a 2 cm side cube was placed in front of the receptive field (shadow regions of the raster). Note in c that opposite but not "mirror-image effects" of equally shaped plastic (before the arrowhead pointing up) and metal (after the arrowhead pointing down) cubes on the "center on" neuron. Spike timing distributions evoked by plastic and metal cube showed a significant difference ($\chi^2$=221, DF=49, p<0.001, 200 spikes taken form left and right boxes in each histogram, Press et al, 1996 ). d) The same unit as illustrated in c, shows sharp variations of the raster responses to the EOD in the presence of "va-et-vien" object's movements. Gray bands encompassing 200 spikes during movements (d) and object absent (e) indicate the data used to construct significantly different histograms f and g ($\chi^2$=91, p<0.001, DF 24). The firing pattern of another "center off" neuron during refuge behavior: h) "head sticking out" condition (local EOD increased); i) control; j) "head fully inside" condition (local EOD reduced). Histograms constructed with 300 spikes each were significantly different ($\chi^2$ tests, DF=49, largest p<0.003 equivalent to 0.01 after Bonferroni's correction).